\documentclass[epj]{svjour}
\usepackage{graphics}
\usepackage{latexsym}
\usepackage[dvips]{graphicx}
\usepackage{exscale}                  
\usepackage[intlimits]{amsmath}       
\usepackage{amsfonts}
\usepackage{amssymb,amscd}
\usepackage[dvips]{epsfig}                   
\usepackage{array}
\usepackage{wrapfig}
\usepackage{graphics}

\begin{document}
\headnote{
    \hspace*{\fill}{\small\sf UNITUE-THEP-23/2002,
                    \hskip .5cm FAU-TP3-02/26} \\[-6pt]
    \hspace*{\fill}{\small\sf http://xxx.lanl.gov/abs/hep-ph/0209366}
                   }
\title{Infrared Exponents and the Running Coupling of Landau gauge QCD and
their Relation to Confinement }
\titlerunning{Infrared Exponents and Running Coupling of QCD}
\author{R.~Alkofer\inst{1}, C.~S.~Fischer\inst{1} and L.~von Smekal\inst{2}
\thanks{Talk given by R.A.\  at the conference
Quark Nuclear Physics 2002.}%
}                     
%
%
\institute{Institute for Theoretical Physics, University of T\"ubingen 
          Auf der Morgenstelle 14, D-72076 T\"ubingen, Germany \and 
	  Institute for Theoretical Physics III, 
             University of Erlangen-N\"urnberg 
             Staudtstr.~7, D-91058 Erlangen, Germany}
\date{September 30, 2002}
\def\makeheadbox{}
%
\abstract{
The infrared behaviour of the gluon and ghost propagators in Landau
gauge QCD is reviewed. The Kugo--Ojima confinement criterion and 
the Gribov--Zwanziger horizon condition result from quite general
properties of the ghost Dyson--Schwinger equation. The numerical
solutions for the gluon and ghost propagators obtained from a truncated
set of Dyson--Schwinger equations provide an explicit example for the
anticipated infrared behaviour. The results are in good agreement with 
corresponding lattice data obtained recently. 
The resulting running coupling approaches a fix point in
the infrared, $\alpha(0) = 8.92/N_c$. Two different 
fits for the scale dependence of the running coupling 
are given and discussed.
} 
\maketitle
%
\section{Aspects of Confinement}
\label{sec:1}
Quarks and gluons, the elementary fields of QCD, are not directly detected in
experiments. Instead, a plethora of hadrons, interpreted as colourless bound
states, are observed. This phenomenon, called {\it confinement}, is still not
properly understood, a clear and undisputable mechanism responsible for this
effect has not been found yet. Moreover it seems not even clear,
at present, whether the phenomenon of confinement is at all compatible
with a description of quark and gluon correlations in terms of local fields
in the usual sense of quantum field theory.

It is interesting to note that the two-point correlations functions of QCD,
the quark, gluon and ghost propagators, might show some signals of the
underlying structures of the theory which are responsible for confinement.
It has been argued that the infrared behaviour of the ghost and the
gluon propagator of ordinary Faddeev--Popov gauge is related to both,
the Kugo--Ojima confinement criterium \cite{Kugo:1979gm}
and the Gribov-Zwanziger horizon condition 
\cite{Gribov:1978wm,Zwanziger:1993qr}.

In the Kugo--Ojima scenario a physical state space that contains
colourless states only is generated if two conditions are fulfilled: 
{\it First} one should not have massless particle poles in transverse gluon
correlations and {\it second} one needs a well-defined, i.e. unbroken,
global colour charge. The second condition can be related
to the behaviour of the ghost propagator in Landau gauge. 
For it to be satisfied, the propagator must be more singular than a
massless  particle  pole in the infrared \cite{Kugo:1995km}.

The Gribov--Zwanziger horizon condition is connected to the
gauge fixing ambiguities in the linear covariant gauge.
Ideally one would eliminate Gribov copies along gauge orbits
by a restriction of the functional integral of the QCD
partition function to the so-called fundamental modular region. This part of
configuration space lies inside the first Gribov region, a convex region in
gauge field space which contains the trivial configuration $A\equiv 0$.
At the boundary of the first Gribov region, the lowest eigenvalue of the 
Faddeev--Popov operator approaches zero. Entropy arguments have
been employed to reason that the infrared modes of the gauge field are  close
to this Gribov horizon \cite{Zwanziger:1993qr}. As the ghost propagator  is the
inverse of the  Fadeev--Popov operator we therefore encounter the presence of
the Gribov horizon in the infrared behaviour of the ghost: The ghost
propagator is required  to be more singular than a simple pole if the
restriction to the Gribov  region is correctly implemented. Furthermore, by the
same entropy arguments, the gluon propagator has to vanish in the infrared
\cite{Zwanziger:1993qr}.


Our framework to investigate the behaviour of the propagators of QCD are the
Dyson--Schwinger equations (DSEs) \cite{Alkofer:2000wg}.  Being complementary
to lattice Monte Carlo simulations which have to deal with finite-volume
effects, DSEs 
allow for analytical investigations of the infrared behaviour of correlation
functions. In Landau gauge we have the particularly simple situation that the
ghost-gluon vertex does not suffer from ultraviolet infinities. Based on this
observation one can use the general structure of the ghost DSE, the properties
of multiplicative renormalizability and the assumption that all involved
Green's functions can be expanded in a power series to show that  the
Kugo--Ojima criterion as well as the Gribov--Zwanziger horizon condition are
satisfied \cite{Watson:2001yv,Lerche:2002ep}. Furthermore, it has been shown
that the infrared behaviour of the ghost and the gluon propagators are uniquely
related: Both fulfill power laws such that  the corresponding powers in the
running coupling (as extracted from the  ghost-gluon vertex) exactly cancel and
one obtains an infrared fix point for the coupling. 

\begin{figure*}
\begin{center}
\epsfig{file=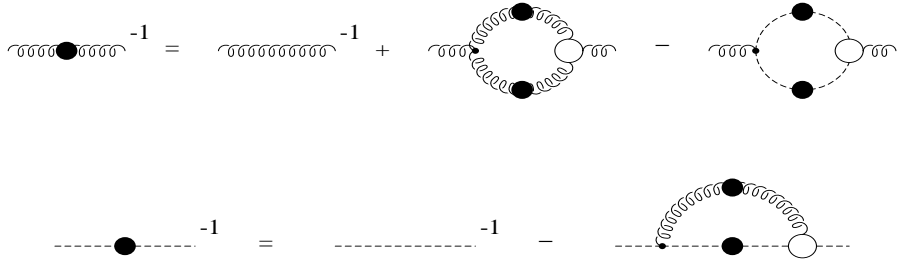,width=12cm,height=36mm}
\end{center}
\caption{Diagrammatic representation of the truncated gluon and ghost 
Dyson--Schwinger equations (DSE). Terms with 
four-gluon vertices have been dismissed. 
Herein, the vertex functions (empty circles) are taken to be
the bare vertices.
}
\label{fig:1}       
\end{figure*}

\section{Infrared exponents and running coupling}
\label{sec:3}

More detailed information on the propagators of Landau gauge QCD can be
obtained from the DSEs once the system is truncated and {\it ansaetze} for the
vertices have been made.  The resulting closed system of equations can be
solved both, analytically in the  infrared and numerically for non-vanishing
momenta. The considerations presented in the previous section suggest that for
small momenta the ghost loop dominates in the gluon DSE. Assuming this
dominance, effects from a wide class of possible dressings for the  ghost-gluon
vertex have been investigated in ref.\ \cite{Lerche:2002ep}  and found  to be
of negligible influence to the qualitative findings.    

Thus, for the purpose of this talk we concentrate on the simplest of these
truncation schemes which has been  developped in detail in refs.\
\cite{Fischer:2002eq,Fischer:2002hn}.  This scheme employs bare three-point
functions and neglects four-gluon vertices, see Fig.~\ref{fig:1}.\footnote{A
coupled system of gluon and ghost DSEs has been studied for the first time in
ref.\ \cite{vonSmekal:1997is}.}  It provides the correct one-loop anomalous
dimensions of the ghost and  gluon dressing functions, $G(k^2)$ and $Z(k^2)$,
respectively,\footnote{These quantities are defined via the gluon and the
ghost propagators via the relations
$D_{\mu \nu}^{\mbox{\tiny Gluon}}(k^2)=
\left(\delta_{\mu \nu} - \frac{k_\mu k_\nu}{k^2}\right)
{Z(k^2)}/{k^2}$ and $D^{\mbox{\tiny Ghost}}(k^2)= - {G(k^2)}/{k^2}$.}
and thus correctly describes the leading logarithmic behaviour of the 
propagators in
the ultraviolet. Furthermore, this scheme reproduces the infrared exponents
found in refs.\ \cite{Lerche:2002ep,Zwanziger:2001kw}:
$Z(k^2) \sim (k^2)^{2\kappa}$ and $G(k^2) \sim (k^2)^{-\kappa}$
with $\kappa \approx 0.595$. These exponents are close to the ones extracted
from lattice calculations \cite{Bonnet:2000kw,Bonnet:2001uh,Langfeld:2001cz}.
Interestingly enough they are also close to the ones obtained in a comparable
truncation scheme in stochastically quantized Landau gauge Yang--Mills theory
\cite{Zwanziger:2002ia}. 

\begin{figure*}
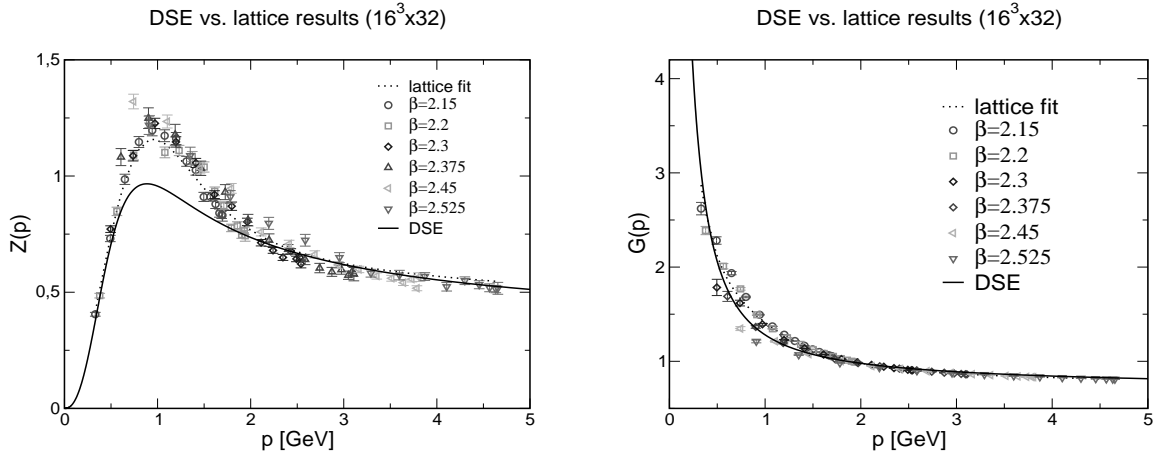

\begin{center}
\epsfig{file=lattice_gluecont.eps,width=7cm,height=6cm}
\hspace{1.0cm}
\epsfig{file=lattice_ghostcont.eps,width=7cm,height=6cm}
\caption{Solutions of the Dyson--Schwinger equations compared to recent 
lattice results for two colours 
\cite{Langfeld:2001cz}}
\end{center}
\label{fig:2}       
\end{figure*}

The numerical solutions are compared to recent lattice calculations 
\cite{Langfeld:2001cz} in Fig.~\ref{fig:2}. Differences mainly
occur for the gluon propagator in the region around the bending point, i.e.\
somewhat below one GeV. These can be attributed
to the omission of the two-loop diagrams in the DSE truncation. Given the
limitations of both methods the qualitative and partly even quantitative 
agreement is remarkable. 

\begin{figure}
\vspace{0.5cm}
\centerline{
\epsfig{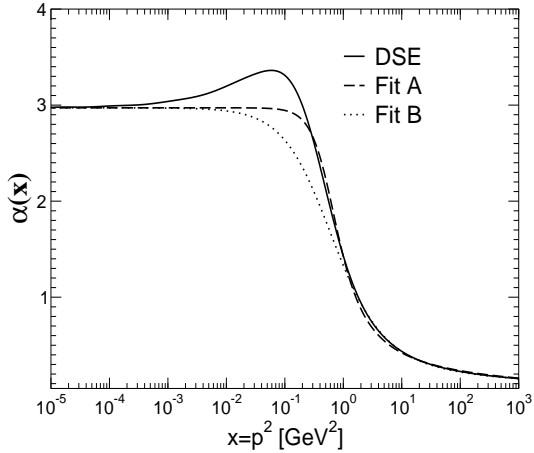}}
\caption{The strong running coupling from the DSEs and the fits 
A and B, c.f.\ eqs.\ (\protect{\ref{fitA},\ref{fitB}}).}
\label{fig:3}       
\end{figure}

The DSE based result for the running coupling can be seen in
Fig.~\ref{fig:3}. The analytically obtained value for the fix point of the
running coupling in the infrared is $\alpha(0) \approx 2.97$ for the gauge
group SU(3) in this truncation scheme. Corrections from possible dressings for
the ghost-gluon vertex have been found to be such that $2.5 < \alpha(0) \le
2.97$ \cite{Lerche:2002ep}. The maximum at non-vanishing momenta
seen in our result for the running coupling results in a multi-valued
beta-function. On the other hand, it appears in a region where the above
comparison to lattice data suggests that our results are least reliable.
(The physical scale has been fixed by requiring the
experimental value $\alpha(M_Z^2=(91.2 \mbox{GeV})^2) = 0.118$.)
We therefore summarize our result for the running coupling
in the monotonic fit functions:\footnote{The $\beta$-function corresponding  
to fit A can be found in ref.~\cite{Alkofer:2002ne}.}  
\begin{eqnarray}
\mbox{Fit A:} \quad
\alpha(x) &=& \frac{\alpha(0)}{\ln(e+a_1 (x/\Lambda^2)^{a_2}+
b_1(x/\Lambda^2)^{b_2})}
\label{fitA}\\
\mbox{Fit B:} \quad
\alpha(x) &=& \frac{1}{a+(x/\Lambda^2)^b} 
\Biggl( a \: \alpha(0) + 
\nonumber \\
&&\left(\frac{1}{\ln(x/\Lambda^2)}
- \frac{1}{x/\Lambda^2 -1}\right)(x/\Lambda^2)^b\Biggr) 
\label{fitB}
\end{eqnarray}
The value $\alpha(0)=2.972=8.915/N_c$ is known from 
the infrared analysis. In both fits the ultraviolet behaviour 
of the solution fixes the scale,  $\Lambda=0.714 \mbox{GeV}$. 
Note that we have employed
a MOM scheme, and thus $\Lambda$ has to be interpreted as
$\Lambda_{MOM}^{N_f=0}$, i.e.\ this scale has the expected magnitude.
Fit A employs the four additional parameters: 
$a_1=1.106$, $a_2=2.324$,
$b_1=0.004$, $b_2=3.169$.
Fit B (which provides a better description in the ultraviolet at the expense
of some deviations at smaller momenta) has only two free parameters:
$a=1.020$, $b=1.052$. 

In summary, we have employed analytical as well numerical studies of the 
gluon and ghost Dyson--Schwinger equations in Landau gauge Yang--Mills theories
to verify the Kugo--Ojima confinement criterion. We have shown that the
resulting infrared behaviour of gluon and ghost propagators, namely a highly
infrared singular ghost and an infrared suppressed gluon propagator, are
related to the Gribov--Zwanziger horizon condition. The solution for these
propagators has then be used to calculate the running coupling for all spacelike
momentum scales.

\section*{Acknowledgements}  
R.A.~thanks the organizers of Quark Nuclear Physics 2002 for the possibility
to participate in this conference.
We are grateful to K.~Langfeld, H.~Reinhardt, D.~Shirkov, 
P.~Watson and D.~Zwanziger for helpful discussions. 

This work has been supported by the DFG under contract Al 279/3-4 and by the
European graduate school T\"ubingen--Basel (DFG contract GRK 683).

\end{document}